\documentclass[conference]{IEEEtran}
\IEEEoverridecommandlockouts
\usepackage{cite}
\usepackage{amsmath,amssymb,amsfonts}
\usepackage{algorithmic}
\usepackage{graphicx}
\usepackage{textcomp}
\usepackage{xcolor}
\def\BibTeX{{\rm B\kern-.05em{\sc i\kern-.025em b}\kern-.08em
    T\kern-.1667em\lower.7ex\hbox{E}\kern-.125emX}}
\begin{document}

\title{Open-source Defect Injection Benchmark Testbed for the Evaluation of Testing}

\author{\IEEEauthorblockN{Miroslav Bures}
\IEEEauthorblockA{\textit{Dept. of Computer Science} \\
\textit{FEE, CTU in Prague}\\
Prague, Czech Republic\\
miroslav.bures@fel.cvut.cz}
\and
\IEEEauthorblockN{Pavel Herout}
\IEEEauthorblockA{\textit{Dept. of Computer Science and Engineering} \\
\textit{University of West Bohemia}\\
Pilsen, Czech Republic\\
herout@kiv.zcu.cz}
\and
\IEEEauthorblockN{Bestoun S. Ahmed}
\IEEEauthorblockA{\textit{Dept. Mathematics \& Comp. Science} \\
\textit{Karlstad University}\\
Karlstad, Sweden \\
bestoun@kau.se}
}

\maketitle

\begin{abstract}

A natural method to evaluate the effectiveness of a testing technique is to measure the defect detection rate when applying the created test cases. Here, real or artificial software defects can be injected into the source code of software. For a more extensive evaluation, injection of artificial defects is usually needed and can be performed via mutation testing using code mutation operators. However, to simulate complex defects arising from a misunderstanding of design specifications, mutation testing might reach its limit in some cases. In this paper, we present an open-source benchmark testbed application that employs a complement method of artificial defect injection. The application is compiled after artificial defects are injected into its source code from predefined building blocks. The majority of the functions and user interface elements are covered by creating front-end-based automated test cases that can be used in experiments. 
\end{abstract}


\section{Introduction}

To evaluate the effectiveness of a testing technique for software systems, various approaches can be employed. A natural and well-known approach to assess the effectiveness of a test suite generated by a testing technique is to measure the defect detection rate when applying a generated test suite to a System Under Test (SUT). As such, an experimental SUT that represents a real-world system containing real defects from the past software development process can be useful here. Alternatively, the mutation testing technique can be applied by introducing artificial defects into the code of an experimental SUT using defined mutation operators \cite{siami2008sufficient, offutt2011mutation}. Additionally, a defect injection technique, which can be considered to be a more general variant, can be employed. In defect injection, defects are introduced into an experimental SUT and various technical possibilities can be used. 

Measuring the defect detection rate can be used to determine the effectiveness of the Combinatorial or Constrained Interaction Testing \cite{nie2011survey,CombConsTBestoun} or Path-based Testing \cite{bures2015pctgen} techniques. As a typical example, one can examine the strength of test cases generated by the Combinatorial Interaction Testing (CIT) algorithm using a mutation testing technique. As an experimental SUT, an open-source software system can be selected. Then, various mutants are created from the source code by a set of mutation operators. Subsequently, the generated test cases are assessed in the experimental SUT and it is determined whether the test case can detect a defect introduced into SUT by a mutation operator. These types of experiments are typically run in multiple series with various sets of mutants and test cases to obtain convincing evidence regarding the effectiveness of the generated test cases \cite{offutt2011mutation}.

\begin{figure}[htbp]
\centerline{\includegraphics[width=7cm]{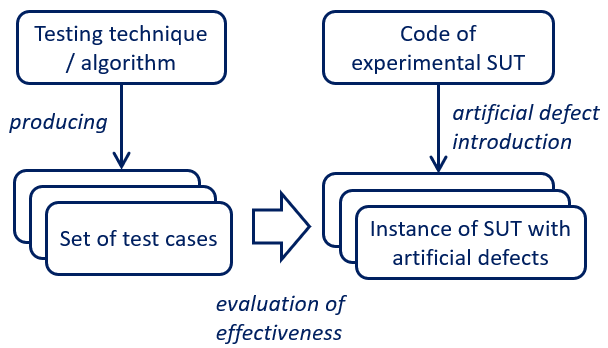}}
\caption{Defect introduction process to an experimental SUT.}
\label{fig:defect_introduction_process}
\end{figure}

The approach can be generalized as illustrated in Figure \ref{fig:defect_introduction_process}.

In this paper, we present a new open-source benchmark testbed to support defect injection testing. The testbed is available to the community and can be used to evaluate various testing techniques. In this approach, we do not insist on defined mutation operators. The goal of the testbed is to provide a complement to the classical mutation testing approach for evaluating the effectiveness of test cases.

In contrast to the established classical code mutation operators, various complex software defects can be introduced into the code, especially defects caused by a misunderstanding of the SUT design specification or requirements during the development process. The practical use case of the presented testbed is to provide researchers with a complementary option to the mutation testing technique to be able to simulate a broader spectrum of possible software defects during experiments. The testbed is, hence, a complement to mutation testing rather a replacement of mutation testing via a defect injection approach. As we show later in Section \ref{sec:background_state_of_the_art}, both approaches have certain advantages and disadvantages. Hence, both approaches can be combined to provide the best objective measurement of the effectiveness of a testing technique.

The rest of this paper is organized as follows. Section \ref{sec:background_state_of_the_art} discusses the background in more depth and analyzes the state of the art. Section \ref{sec:testbed_description} describes the presented testbed from various viewpoints, including the system scope, implementation details, available automated tests, mechanism of insertion of artificial defects and process of evaluating the effectiveness of the examined testing techniques. Section \ref{sec:discussion_and_possible_limits} discusses the presented concept and also analyzes its possible limits. The last section concludes the paper.


\section{Background and State of The Art}
\label{sec:background_state_of_the_art}

As mentioned previously, a common practice to evaluate a set of test cases generated by an algorithm is to assess the defect detection rate of the test cases in an experimental SUT that contains defects. In this general approach, several aspects have to be maintained to give the technical possibility of conducting a well-defined and objective experiment. The following bullet-points address three common aspects in this direction:

\begin{enumerate}
\item The defects presented in the experimental SUT simulate real defects in software projects.
\item It is possible to create a set of various instances of an experimental SUT with different sets of injected defects to examine the testing technique for a reliable sample of situations.
\item The experimental SUT has to support effective automated evaluation of the examined test cases. Hence, the experiments can be repeated with different sets of defects in an effective manner to assess more extensive sets of situations.
\end{enumerate}

Table \ref{tab:comparison_of_defect_introduction_techniques} presents an analysis of these aspects for three possibilities of artificial defect introduction within an experimental SUT. These possibilities are as follows: (1) using real project defects, (2) mutation testing, and (3) defect injection. Defect injection, here, is a generalized method in which we do not employ standard source code mutation operators. In fact, it is difficult to reach a clear understanding of an objective approach from all three discussed options when considering all the advantages and disadvantages presented in Table \ref{tab:comparison_of_defect_introduction_techniques}. Instead, it is worthwhile to consider a combination of the presented approaches to increase the reliability of the experiments.

\begin{table*}
\caption{Brief comparison of artificial defect introduction types to an experimental SUT}\label{tab:comparison_of_defect_introduction_techniques}
\begin{centering}
\begin{tabular}{|p{1.8cm}|p{4.4cm}|p{4.4cm}|p{4.5cm}|}
\hline
\textbf{Discussed}&\multicolumn{3}{|c|}{\textbf{Defect introduction method}} \\
\cline{2-4} 
\textbf{aspects} & \textbf{\textit{Historic defects}}& \textbf{\textit{Mutation testing}}& \textbf{\textit{Defect injection}} \\
\hline
The objectivity of the defects$^{\mathrm{a}}$ & The defects correspond to a real software project; however, the used sample of defects can be limited, which can restrict the objectivity of the experiment to only one particular experience-based case. &
Various combinations of mutation operators can be selected. This approach allows the flexible mixing of various defects made by the programmer. More complex defects caused by a misunderstanding of the specification can be simulated by a set of mutation operators. 
& More complex simulated defects are not limited to a defined set of mutation operators. Additionally, it might be difficult to prove that an artificially elaborated defect is likely to occur in the real software development process. 
\\
\hline
Ease to create instances$^{\mathrm{b}}$  & In some cases, creating multiple instances might be challenging, as there are a limited number of defects from the past software development process.
& Technically, creating new mutants is straightforward, and the number of various created SUT instances is practically unlimited. & 
If a set of artificially elaborated defects is limited, then the possible number of instances of experimental SUTs that can be configured is limited.

\\
\hline
Test automation coverage$^{\mathrm{c}}$&  \multicolumn{3}{|p{13.8cm}|}{Test automation options are not influenced by a particular defect introduction method; automated testability is rather influenced by the structure and coding standards employed in an experimental SUT}  \\
\hline
\multicolumn{4}{l}{$^{\mathrm{a}}$How the introduced defects are realistic in comparison to real current software development process}\\

\multicolumn{4}{l}{$^{\mathrm{b}}$How easy is it to create an extensive set of various configurations of an experimental SUT with different inserted defects}\\

\multicolumn{4}{l}{$^{\mathrm{c}}$How easy is it to cover an experimental SUT by automated tests that help to evaluate whether the examined test scenarios detect an inserted defect}

\end{tabular}

\end{centering}
\end{table*}

Among the discussed approaches, mutation testing can be considered to be the most established approach, originating in the late 70s \cite{demillo1979program}. On the technical level, this approach depends on a particular programming language. However, code mutations have been performed for major programming languages. As an example, the Mujava system \cite{ma2005mujava} is used for the Java programming language and MuCPP \cite{delgado2017assessment} is used for C++. Here, for a particular program code mutation, a set of established operators is defined \cite{siami2008sufficient, offutt2011mutation}. While these operators are useful, there are concerns in the literature about the relation of the code mutants to real software defects and types of software defects that are difficult to express using various mutants \cite{papadakis2019mutation,gopinath2014mutations}. To overcome this problem, various approaches have been considered in the literature -- for instance, the construction of more complex mutants \cite{papadakis2019mutation}. However, the mutation testing approach might still meet its limit when trying to insert certain types of complex defects that may be caused by a misunderstanding of the design specification \cite{gopinath2014mutations}. Generally, the similarity of mutants to real defects varies in empirical experiments \cite{gopinath2014mutations,andrews2005mutation}.


The defect injection method can be seen as a more general method than mutation testing to insert artificial defects into experimental software. In this process, various techniques at any software level can be used to insert defects, e.g., \cite{cotroneo2012experimental,kooli2014survey}. As an alternative to mutation testing and artificial defect injection, a number of experiments have also been conducted using real defects from past software projects, e.g., \cite{bures2018tapir}. Here, comparing those different approaches is challenging because the objectivity of the experiment in which we evaluate the effectiveness of the testing techniques strongly depends on the testing technique, the characteristics of the software used as a benchmark, and how realistic the inserted defects are compared to real defects. Moreover, the characteristics of the software defects might also change with changes in the development styles, the usage of integrated development environments and the best practices of programming. To this end, in this paper, we suggest applying a benchmark testbed as a complement to the mutation testing approach.

\section{Testbed Description}
\label{sec:testbed_description}

To create a benchmark testbed for the evaluation of the effectiveness of the test technique, we designed and implemented the University Information System Testbed (TbUIS)\footnote{https://projects.kiv.zcu.cz/tbuis/}. The testbed is an open-source testbed that can be used to evaluate any test technique. The TbUIS system is a three-layered web application that uses a relational database as a persistent data storage and object-relational mapping (ORM) layer.

The system supports the artificial defect injection approach, as discussed in Table \ref{tab:comparison_of_defect_introduction_techniques} (column \textit{Defect injection}). A special module allows the creation of defect clones of the system by introducing defects from a catalog of predefined defect types as well as creating customized artificial defects to be inserted into the SUT. As a demonstration and for a quick start for experiments\color{black}, a set of 28 already assembled defect clones are available for testbed users.


In this section, we describe the following aspects of the TbUIS: (A) the scope of the system and its use cases, (B) the implementation and technical details, (C) the available automated tests to be employed in the experiments, (D) the mechanism for introducing artificial defects in the system and (E) the test case effectiveness evaluation process, including the logging mechanism used in the evaluations.

\subsection{Scope and Use Cases of the TbUIS}
\label{sec:scope_and_use_cases}

The TbUIS is a fictional university study information system that supports a study agenda related to students’ enrolment in courses, management of exams and related processes. The standard actors of the system are students and lecturers.  

The whole system can be summarized into 21 general, high-level use cases. Five use cases are related to a user who is not logged in. Another two use cases are common for lecturers and students and cover the login mechanism and user settings. The students’ part is then defined by five use cases and the lecturers’ part by nine separate use cases.

\begin{figure}
\centerline{\includegraphics[width=8.5cm]{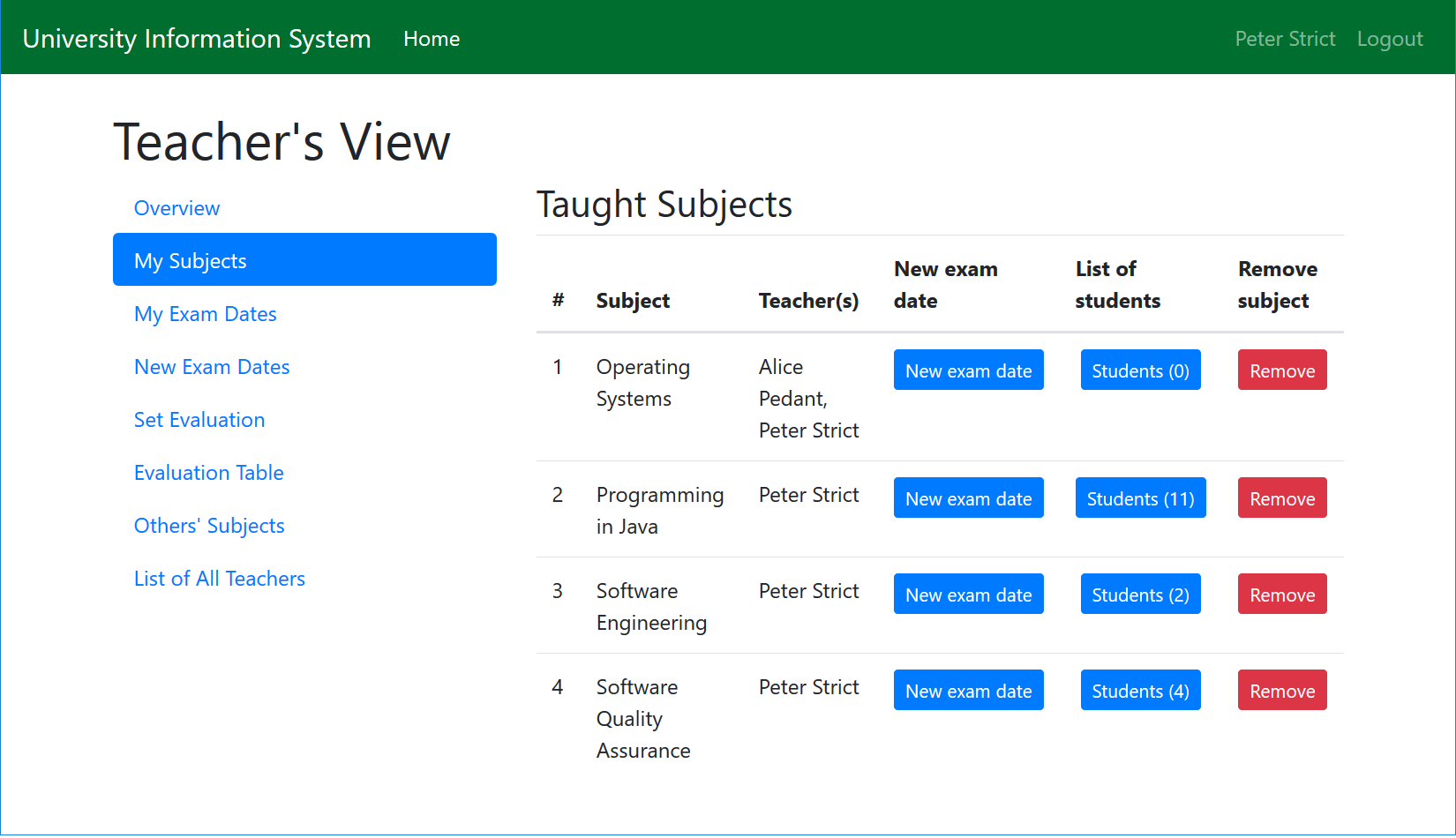}}
\caption{Example of TbUIS user interface---lecturer's view.}
\label{fig:ui_sample}
\end{figure}

The graphical layout of the user interface (UI) of the system is kept relatively simple and compact, considering the goal of the system, which is to evaluate testing techniques as well as the need to cover the application by reliable front-end (FE) based automated tests to support this process (introduced later in Section \ref{sec:automated_tests}). An example of the TbUIS user interface is presented in Figure \ref{fig:ui_sample}.

Regarding the process flow, the possible states and functions of the system are documented in the UML Activity Diagram schema in the Oxygen\footnote{http://still.felk.cvut.cz/oxygen/} \cite{bures2015pctgen} application and are available in the Oxygen project format as well as the SVG graphical format. This model of the current version of the TbUIS is composed of 119 different states and 164 transitions among them.

\subsection{Implementation and Technical Details}

Technically, the TbUIS is a layered web-based application implemented in J2EE with Java Server Pages (JSP) and Spring. As the ORM layer, Hibernate is used. For the implementation of the UI, Bootstrap is used.

In the user interface of the TbUIS, all the common basic types of control for web elements (e.g., menus, buttons, check boxes, selections, modal windows, etc.) are used. Each element (including rows in tables) has its own unique ID attribute to ease the creation of FE-based functional automated tests.

The extent of the TbUIS source code is documented in Table \ref{tab:source_code_size}. Here, the number of source files, size of source code files in kilobytes and number of lines of code (LOC) are presented separately for back-end code in Java as well as for UI code in JSP. Unit tests as well as functional automated tests are not included in these statistics.

\begin{table}
\begin{center}
\caption{Size of TbUIS source code}
\begin{tabular}{|r|c|c|c|}\hline
     & Number of files & Size of files [KB] & LOC \\ \hline\hline
     Java & 87 & 340 & 8550 \\ \hline
     JSP & 18 & 94 & 1550 \\ \hline\hline
     total & 105 & 434 & 10100 \\ \hline
\end{tabular}
\label{tab:source_code_size}
\end{center}
\end{table}

Any important activity in the TbUIS testbed is reported in detailed application logs implemented by Log4J2. Because of the logging framework configuration options, the user can customize the level of detail and the output stream of the log. The application logs are also extended by the activation information of the inserted artificial defects (further discussed in Section \ref{sec:artificial_defects}) and can be paired with the logs of available functional automated tests (further discussed in Section \ref{sec:automated_tests}).

\subsection{Automated tests}
\label{sec:automated_tests}

TbUIS is strongly covered by various types of automated tests that have the following two goals:

\begin{enumerate}
\item To ensure that the system (before the introduction of controlled artificial defects used to evaluate the effectiveness of testing techniques) is largely free of other defects and

\item To support the process of evaluating the effectiveness of the testing techniques by executing the defined test cases that are to be examined in the system via automated tests,
\end{enumerate}

Two types of tests are available as extra modules for the TbUIS testbed:

\begin{enumerate}
    \item \textbf{Unit tests} implemented in the JUnit framework, which test individual methods of the system and the basic sequences of methods calls on the technical level.
    \item \textbf{FE-based functional tests}, which simulate users’ tests accessing the system UI. These tests are written in Java with the Selenium Web Driver API, currently version 3.141.59. The tests are structured using the PageObject pattern, which significantly decreases their maintenance and allows future extensions of the test set, as independently verified \cite{bures2015model}.
 \end{enumerate}

Regarding the coverage level, in the current version of the TbUIS, the line coverage of the available unit tests is greater than 85\%.

The FE-based functional tests cover all of the processes, as documented in the process flow schema created in the Oxygen application (introduced above in Section \ref{sec:scope_and_use_cases}).

To determine the expected test results of the FE-based functional tests, the Oracle module is implemented and is thoroughly tested using a special set of unit tests.

Table \ref{tab:source_codes_of_tests} provides insight into the extent of the implemented automated tests. The number of source code files, their size in kilobytes and the number of lines of code (LOC) are presented. The FE-based functional tests for TbUIS employ several modules of reusable objects and support code (in Table \ref{tab:source_codes_of_tests} denoted as \textit{Shared libraries for FE-based functional tests}). These modules are also covered by their own set of unit tests.

\begin{table}
\begin{center}
\caption{Extent of source code of automated tests}
\begin{tabular}{|p{2.5cm}|c|c|c|}\hline
     & Number of files & Size of files [KB] & LOC \\ \hline\hline
     Unit tests for TbUIS code& \hfill{}\lower1.8mm\hbox{33}\hspace*{7mm} & \hfill{}\lower1.8mm\hbox{277}\hspace*{8mm} & \hfill{}\lower1.8mm\hbox{6945}\hspace*{0.5mm} \\ \hline
     Shared libraries for FE-based functional tests & \hfill{}\lower3.0mm\hbox{101}\hspace*{7mm} & \hfill{}\lower3.0mm\hbox{452}\hspace*{8mm} & \hfill{}\lower3.0mm\hbox{14707}\hspace*{0.5mm} \\ \hline
     Unit tests for shared libraries for FE-based functional tests & \hfill{}\lower3.0mm\hbox{30}\hspace*{7mm} & \hfill{}\lower3.0mm\hbox{97}\hspace*{8mm} & \hfill{}\lower3.0mm\hbox{2649}\hspace*{0.5mm} \\ \hline
     FE-based functional tests for TbUIS & \hfill{}\lower1.8mm\hbox{81}\hspace*{7mm} & \hfill{}\lower1.8mm\hbox{248}\hspace*{8mm} & \hfill{}\lower1.8mm\hbox{7530}\hspace*{0.5mm} \\ \hline\hline
     total & \hfill{}245\hspace*{7mm} & \hfill{}1024\hspace*{8mm} & \hfill{}31831\hspace*{0.5mm} \\ \hline
\end{tabular}
\label{tab:source_codes_of_tests}
\end{center}
\end{table}
Compared to size of the source code of the TbUIS (see Table \ref{tab:source_code_size}), the extent of the automated tests measured in terms of LOC is approximately three times higher.

The FE-based functional automated tests are divided into several types, covering various technical and user aspects of the TbUIS:

\begin{itemize}
\item \textbf{Atomic tests} that are verifying if elements of application UI are correctly rendered and filled with correct data
\item \textbf{Process tests} that are exercising individual processes in the TbUIS (e.g. enrolling a course or assigning a grade to the student)
\item \textbf{Negative tests} that are testing boundary conditions and correct handling of wrong input data
\end{itemize}

Atomic types of tests are also orchestrated as parts of the process tests. The test scripts are organized into building blocks that allow the automated composition of an automated end-to-end test via a defined path-based test scenario (the details are presented in Section \ref{sec:test_case_effectiveness_evaluation}).

The numbers of tests in the individual categories with their numbers of asserts and average runtime are presented in Table \ref{tab:testing_scale}. The runtimes were measured using the following configuration: Intel i5 1.6 GHz, 16 GB RAM, MS Windows 10pro operating system, Apache Tomcat 9.0 application server and MySQL database. The database and web and application servers were installed on the same workstation, and the automated tests were run on the same computer.

\begin{table}
\begin{center}
\caption{Types of FE-based functional automated tests}
\begin{tabular}{|l|p{1.2cm}|p{1.2cm}|p{1.2cm}|}\hline
     & Number of tests & Number of asserts & Elapsed time [sec] \\ \hline\hline
     Atomic tests & \hfill{}890\hspace*{4mm} & \hfill{}2702\hspace*{4mm} & \hfill{}780\hspace*{4mm} \\ \hline
     Process tests & \hfill{}64\hspace*{4mm} & \hfill{}2351\hspace*{4mm} & \hfill{}1477\hspace*{4mm} \\ \hline
     Negative tests & \hfill{}29\hspace*{4mm} & \hfill{}52\hspace*{4mm} & \hfill{}50\hspace*{4mm} \\ \hline\hline
     total & \hfill{}983\hspace*{4mm} & \hfill{}5105\hspace*{4mm} & \hfill{}2307\hspace*{4mm} \\ \hline
\end{tabular}
\label{tab:testing_scale}
\end{center}
\end{table}

The automated atomic tests cover 100\% of all active and passive elements composing the user interface of the TbUIS. As active elements, we consider user control elements (e.g., text boxes, drop-down menus, links, etc.) and fields that display data loaded from the database or are taken from the runtime memory of the application. Each of the active elements is tested at least by one atomic test.

FE-based automated functional tests can be easily run from a special application, TestRunner, which provides its own user interface in which particular tests to run can be selected. The TestRunner application can be downloaded from the project web page.

The extent of the building blocks of the FE-based automated functional tests introduced in this section allows the effective composition of automated tests for the path-based test scenarios to be evaluated in the testbed. The relevant part of these blocks can also be used to evaluate the combinatorial or constrained interaction testing test sets.

\subsection{Introduction of Artificial Defects}
\label{sec:artificial_defects}

Artificial defects are introduced into the TbUIS by the error seeder module, which conducts the following process:
\begin{enumerate}
\item The error seeder takes the baseline TbUIS code, which is considered free of defects (which is verified by the thorough automated tests introduced in Section \ref{sec:automated_tests}). 
\item Based on the artificial defect specification, the error seeder assembles the source code of a defect clone of the TbUIS.
\item Then, the defect clone of the TbUIS system is built and deployed to a testing environment.
\end{enumerate}

Artificial defect specification defines a set of artificial defects that are to be introduced into the TbUIS code. Predefined catalogue defect types are available as well as the possibility to define custom artificial defects. The catalogue of defect types is available on the project web page.

Each artificial defect inserted into the TbUIS code is accompanied by a logging mechanism that records information if and when the defect has been activated by a test. The main purpose of this information is to support the evaluation of the effectiveness of the testing techniques. The defect activation logs can be paired with the logs of available automated tests to give reliable sources of information, which artificial defect were detected by which test cases.

In the current version of the TbUIS testbeds, a set of 27 artificial defects of various types from the above catalogue is available for initial experiments and are accompanied by detailed information making their application easy\footnote{https://projects.kiv.zcu.cz/tbuis/web/page/download}.

As mentioned above, for further experiments and to evaluate the effectiveness of the testing techniques, more various defect clones can be created and compiled from available artificial defects, and also, based on the well-documented examples in the source code, the user can implement their own artificial defects.

\subsection{Test Case Effectiveness Evaluation Process}
\label{sec:test_case_effectiveness_evaluation}

As introduced above, in principle, the effectiveness of various testing techniques can be evaluated in the TbUIS testbed. In the following section, we focus on two major representatives, path-based techniques and combinatorial/constrained interaction testing techniques.

\begin{figure}
\centerline{\includegraphics[width=9cm]{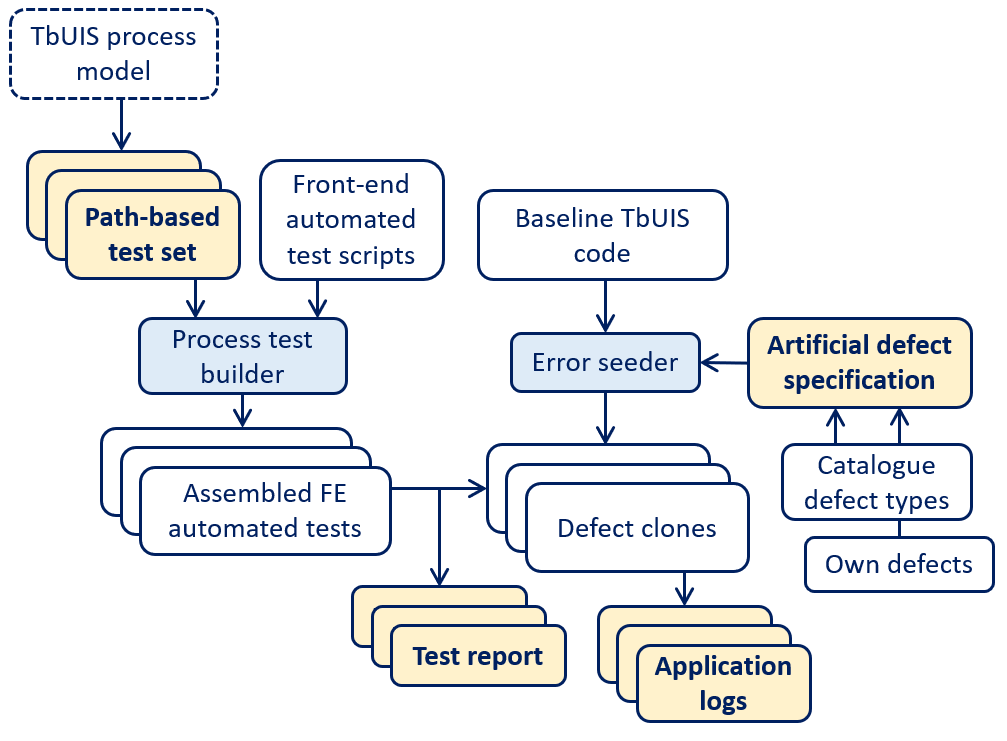}}
\caption{TbUIS parts for evaluation of path-based testing techniques.}
\label{fig:testbed_parts_paths}
\end{figure}

The parts of the TbUIS testbed related to the evaluation of path-based testing techniques are summarized in Figure \ref{fig:testbed_parts_paths}. The inputs and outputs of the process are depicted by yellow boxes.

The input to the process is a \textit{path-based test set}, whose effectiveness is going to be evaluated. The test cases in this test set have to correspond to an available \textit{TbUIS process model} (unless we intentionally created invalid paths-based test cases in the experiment). Using predefined building blocks from the \textit{FE-based functional automated test scripts} (introduced in Section \ref{sec:automated_tests}), the \textit{process test builder} chains these building blocks as instructed by the input path-based test cases to produce \textit{assembled FE automated tests}, which represent individual path-based test cases. For each of the path-based test cases at the input, a corresponding automated FE test is created.

The second input of the process is \textit{artificial defect specification}, which can be created via predefined \textit{catalogue defect types} or \textit{own defects} defined in the SUT. To create an \textit{defect clone} of the TbUIS with the specified defects, \textit{Error seeder} takes the specification of the defects and inserts them into the code of the Baseline UIS specification. Then, the defect clone is built as a running system instance.

At this stage, experimental evaluation of the path-based test set can be performed (an example is provided on the project web pages). Automated FE tests corresponding to the input path-based test cases are run in the defect clone, and the results are reported to the \textit{test report}, which can be evaluated. The information from the test report can also be paired with detailed \textit{application logs} to obtain more context information about the activated defects.

The schema for evaluating combinatorial or constrained testing test cases slightly differs, but the general principle remains the same.

In this type of evaluation, we do not compile FE automated tests to correspond to path-based test cases; instead, we can use
\begin{enumerate}
\item available automated FE-based functional tests covering all active elements and processes in the TbUIS, 
\item available unit tests available together with the TbUIS code, or
\item combinations of both types of tests (the automated tests available to the TbUIS were introduced in Section \ref{sec:automated_tests}).
\end{enumerate}

Input data combinations to be exercised in the testbed can be entered into the available automated tests via the standardized DataProvider interface of the JUnit framework.

\section{Discussion and Possible Limits}
\label{sec:discussion_and_possible_limits}

Like other alternative artificial defect introduction approaches discussed in this paper, namely, using real defects from a previous software project and code mutation, the approach taken in the proposed testbed has certain advantages and disadvantages. We summarize these advantages and disadvantages in this section.

Regarding the possible complexity of the artificial defects introduced into an experimental SUT, the proposed approach does not limit an artificial defect to a set of mutation operators or a conditionally switched block of code. Instead, the defect clone can be built with the changes made in several different places in the source code, which allows high flexibility in simulating complex defects.

Concern whether the introduced defects represent typical defects that are being made during real software projects can be raised. This responsibility in experiments is up to the researchers and testing practitioners. Typical defects might vary between various software architectures, development styles, programming languages, business domains, and even decades when the empirical observations are made. Hence, the testbed provides a general possibility to create different types of defects and defect clones, and the decision is up to the testbed user.

In the proposed concept, the artificial defects are selected from a pre-defined set, which might limit the generalization of experiment results. This potential limit can be solved by the addition of more artificial defects as well as the correct interpretation of the results of the experiments.

Also, certain defects might be easier to detect than other defects, which may impact the results of the experiments \cite{papadakis2016threats}. However, this concern can be raised generally for any defect injection technique and shall be mitigated by correct interpretation of the results of the experiments.

Another concern is that the system is artificially created; however, the use cases and processes in the SUT are similar to real-world study information systems. The more important factor here is the selection of artificial defects that are representative of real-world projects. In the presented testbed, this selection is enabled by the possible introduction of more complex defects via the described mechanism of the defect clones.

Also, the size of the TbUIS system might limit its potential applicability as a benchmark for larger software systems. We are going to mitigate this concern by further evolution and extensions of the TbUIS.

\section{Conclusion}

In evaluating the effectiveness of testing techniques based on the measurement of the defect number that the test cases produced by these techniques detect in an experimental system, the established mutation testing approach can be accompanied by an alternative allowing the insertion of more complex defects caused by a misunderstanding of the design specification or other causes. We describe such an alternative in this paper: the presented TbUIS testbed, which is available as an open-source application and comprises a fictional university information system. The TbUIS testbed gives its user a mechanism to introduce artificial defects, including those from a predefined catalogue of possible defects, an extensive set of unit and FE-based functional automated tests, which can be used to examine test cases in the system, and a logging mechanism, which allows the collection of the data regarding which defects were activated by the examined test cases. Together with a good level of code and system documentation, the open structure of the TbUIS testbed eases its employment as a benchmark system to be used in the evaluation of path-based and combinatorial/constrained interaction testing techniques.

\section*{Acknowledgment}

This work was supported by the European structural and investment funds (ESIF) project CZ.02.1.01/0.0/0.0/17\_048/0007267 (InteCom)---Intelligent Components of Advanced Technologies for the Pilsen metropolitan area. Work package WP1.3: Methods and processes for control software safety assurance. The authors acknowledge the support of the OP VVV funded project CZ.02.1.01/0.0/0.0/16\_019/0000765 “Research Center for Informatics”.

\bibliographystyle{IEEEtran}
\bibliography{references}

\begin{thebibliography}{10}
\providecommand{\url}[1]{#1}
\csname url@samestyle\endcsname
\providecommand{\newblock}{\relax}
\providecommand{\bibinfo}[2]{#2}
\providecommand{\BIBentrySTDinterwordspacing}{\spaceskip=0pt\relax}
\providecommand{\BIBentryALTinterwordstretchfactor}{4}
\providecommand{\BIBentryALTinterwordspacing}{\spaceskip=\fontdimen2\font plus
\BIBentryALTinterwordstretchfactor\fontdimen3\font minus
  \fontdimen4\font\relax}
\providecommand{\BIBforeignlanguage}[2]{{%
\expandafter\ifx\csname l@#1\endcsname\relax
\typeout{** WARNING: IEEEtran.bst: No hyphenation pattern has been}%
\typeout{** loaded for the language `#1'. Using the pattern for}%
\typeout{** the default language instead.}%
\else
\language=\csname l@#1\endcsname
\fi
#2}}
\providecommand{\BIBdecl}{\relax}
\BIBdecl

\bibitem{siami2008sufficient}
A.~Siami~Namin, J.~H. Andrews, and D.~J. Murdoch, ``Sufficient mutation
  operators for measuring test effectiveness,'' in \emph{Proceedings of the
  30th international conference on Software engineering}.\hskip 1em plus 0.5em
  minus 0.4em\relax ACM, 2008, pp. 351--360.

\bibitem{offutt2011mutation}
J.~Offutt, ``A mutation carol: Past, present and future,'' \emph{Information
  and Software Technology}, vol.~53, no.~10, pp. 1098--1107, 2011.

\bibitem{nie2011survey}
C.~Nie and H.~Leung, ``A survey of combinatorial testing,'' \emph{ACM Computing
  Surveys (CSUR)}, vol.~43, no.~2, p.~11, 2011.

\bibitem{CombConsTBestoun}
B.~S. Ahmed, K.~Z. Zamli, W.~Afzal, and M.~{Bures}, ``Constrained interaction
  testing: A systematic literature study,'' \emph{IEEE Access}, vol.~5, pp.
  25\,706--25\,730, 2017.

\bibitem{bures2015pctgen}
M.~Bures, ``Pctgen: Automated generation of test cases for application
  workflows,'' in \emph{New Contributions in Information Systems and
  Technologies}.\hskip 1em plus 0.5em minus 0.4em\relax Cham: Springer
  International Publishing, 2015, pp. 789--794.

\bibitem{demillo1979program}
R.~A. DeMillo, R.~J. Lipton, and F.~G. Sayward, ``Program mutation: A new
  approach to program testing,'' \emph{Infotech State of the Art Report,
  Software Testing}, vol.~2, no. 1979, pp. 107--126, 1979.

\bibitem{ma2005mujava}
Y.-S. Ma, J.~Offutt, and Y.~R. Kwon, ``Mujava: an automated class mutation
  system,'' \emph{Software Testing, Verification and Reliability}, vol.~15,
  no.~2, pp. 97--133, 2005.

\bibitem{delgado2017assessment}
P.~Delgado-P{\'e}rez, I.~Medina-Bulo, F.~Palomo-Lozano,
  A.~Garc{\'\i}a-Dom{\'\i}nguez, and J.~J. Dom{\'\i}nguez-Jim{\'e}nez,
  ``Assessment of class mutation operators for c++ with the mucpp mutation
  system,'' \emph{Information and Software Technology}, vol.~81, pp. 169--184,
  2017.

\bibitem{papadakis2019mutation}
M.~Papadakis, M.~Kintis, J.~Zhang, Y.~Jia, Y.~Le~Traon, and M.~Harman,
  ``Mutation testing advances: an analysis and survey,'' in \emph{Advances in
  Computers}.\hskip 1em plus 0.5em minus 0.4em\relax Elsevier, 2019, vol. 112,
  pp. 275--378.

\bibitem{gopinath2014mutations}
R.~Gopinath, C.~Jensen, and A.~Groce, ``Mutations: How close are they to real
  faults?'' in \emph{2014 IEEE 25th International Symposium on Software
  Reliability Engineering}.\hskip 1em plus 0.5em minus 0.4em\relax IEEE, 2014,
  pp. 189--200.

\bibitem{andrews2005mutation}
J.~H. Andrews, L.~C. Briand, and Y.~Labiche, ``Is mutation an appropriate tool
  for testing experiments?'' in \emph{Proceedings of the 27th international
  conference on Software engineering}.\hskip 1em plus 0.5em minus 0.4em\relax
  ACM, 2005, pp. 402--411.

\bibitem{cotroneo2012experimental}
D.~Cotroneo, A.~Lanzaro, R.~Natella, and R.~Barbosa, ``Experimental analysis of
  binary-level software fault injection in complex software,'' in \emph{2012
  Ninth European Dependable Computing Conference}.\hskip 1em plus 0.5em minus
  0.4em\relax IEEE, 2012, pp. 162--172.

\bibitem{kooli2014survey}
M.~Kooli and G.~Di~Natale, ``A survey on simulation-based fault injection tools
  for complex systems,'' in \emph{2014 9th IEEE International Conference On
  Design \& Technology of Integrated Systems In Nanoscale Era (DTIS)}.\hskip
  1em plus 0.5em minus 0.4em\relax IEEE, 2014, pp. 1--6.

\bibitem{bures2018tapir}
M.~Bures, K.~Frajtak, and B.~S. Ahmed, ``Tapir: Automation support of
  exploratory testing using model reconstruction of the system under test,''
  \emph{IEEE Transactions on Reliability}, vol.~67, no.~2, pp. 557--580, 2018.

\bibitem{bures2015model}
M.~Bures, ``Model for evaluation and cost estimations of the automated testing
  architecture,'' in \emph{New Contributions in Information Systems and
  Technologies}.\hskip 1em plus 0.5em minus 0.4em\relax Cham: Springer
  International Publishing, 2015, pp. 781--787.

\bibitem{papadakis2016threats}
M.~Papadakis, C.~Henard, M.~Harman, Y.~Jia, and Y.~Le~Traon, ``Threats to the
  validity of mutation-based test assessment,'' in \emph{Proceedings of the
  25th International Symposium on Software Testing and Analysis}, 2016, pp.
  354--365.

\end{thebibliography}

\end{document}